\documentclass[prb,twocolumn,showpacs,superscriptaddress,preprintnumbers,floatfix,amsmath,amssymb]{revtex4-1}

\usepackage{amssymb}
\usepackage{graphicx}
\usepackage{dcolumn}
\usepackage{bm}
\usepackage{color}
\begin{document}

\title{Magnetism and superconductivity in Eu(Fe$_{1-x}$Ni$_{x}$)As$_2$ ($x$ = 0, 0.04)}

\author{Ya-Bin Liu}
\affiliation{Department of Physics, Zhejiang University, Hangzhou 310027, China}
\author{Yi Liu}
\affiliation{Department of Physics, Zhejiang University, Hangzhou 310027, China}
\author{Wen-He Jiao}
\affiliation{Department of Physics, Zhejiang University of Science and Technology, Hangzhou 310023, China}
\author{Zhi Ren}
\affiliation{Institute for Natural Sciences, Westlake Institute for Advanced Study, Hangzhou 310013, China}
\author{Guang-Han Cao}\email[]{ghcao@zju.edu.cn}
\affiliation{Department of Physics, Zhejiang University, Hangzhou
310027, China}
\affiliation{Collaborative Innovation Centre of Advanced Microstructures, Nanjing University, Nanjing 210093, China}
\affiliation{State Key Lab of Silicon Materials, Zhejiang University, Hangzhou 310027, China}

\date{\today}

\begin{abstract}
We report Eu-local-spin magnetism and Ni-doping-induced superconductivity (SC) in a 112-type ferroarsenide system Eu(Fe$_{1-x}$Ni$_{x}$)As$_2$. The non-doped EuFeAs$_2$ exhibits two primary magnetic transitions at $\sim$100 and $\sim$ 40 K, probably associated with a spin-density-wave (SDW) transition and an antiferromagnetic ordering in the Fe and Eu sublattices, respectively. Two additional successive transitions possibly related to Eu-spin modulations appear at 15.5 and 6.5 K. For the Ni-doped sample with $x$ = 0.04, the SDW transition disappears, and SC emerges at $T_\mathrm{c}$ = 17.5 K. The Eu-spin ordering remains at around 40 K, followed by the possible reentrant magnetic modulations with enhanced spin canting. Consequently, SC coexists with a weak spontaneous magnetization below 6.2 K in Eu(Fe$_{0.96}$Ni$_{0.04}$)As$_2$, which provides a complementary playground for the study of the interplay between SC and magnetism.
\end{abstract}


\maketitle

\section{Introduction}\label{sec:1}

The discovery of superconductivity (SC) in LaFeAsO$_{1-x}$F$_x$~\cite{hosono08} leads to the successive findings of dozens of iron-based superconductors with various crystal structures~\cite{johrendt08,wmk08,jcq08,whh09,cxl10,kakiya11,cava11,syl12,
karpinski13,nohara13,ogino14,cxh15,iyo16,wzc16}. Among these superconductors, Eu-containing iron pnictides stand out owing to the Eu-moment \emph{ferromagnetic} ordering in the superconducting state~\cite{rz09a,nandi14,MFM17,jwh17,dressel17}. The prototype example is the EuFe$_2$As$_2$ system, a unique 122-type iron pnictide in which the Eu$^{2+}$ sublattice in between the superconductively active Fe$_2$As$_2$ layers carries a local spin with $S=7/2$~\cite{raffius93,rz08,js09a}. Upon a certain chemical doping~\cite{jeevan08,rz09a,js09b,rza10,jwh11,jwh13} or with applying physical pressures~\cite{miclea09,dressel17}, the system may change into a ferromagnetic superconductor which simultaneously exhibits SC and Eu-spin ferromagnetism below their transition temperatures. Recently, Eu-containing 1144-type compounds $A$EuFe$_4$As$_4$ ($A$ = Rb, Cs) were discovered~\cite{kawashima16,ly16a,ly16b}. These materials can be viewed as an intergrowth of EuFe$_2$As$_2$ and $A$Fe$_2$As$_2$ in which the Eu-atomic layer remains intact. The Eu subsystem shows ferromagnetic behavior below 15 K, coexisting with SC that emerges at a higher temperature of about 35 K~\cite{ly16a,ly16b,stadnik18a,stadnik18b}.

Very recently, a new Eu-containing compound EuFeAs$_2$ was reported~\cite{rza17}. Its crystal structure is analogous to the 112-type iron-pnictide superconductors (Ca,$Ln$)FeAs$_2$ ($Ln$ = La~\cite{nohara13} and Pr~\cite{ogino14}), consisting of alternately stacked Fe$_2$As$_2$ and As-zigzag-chain layers. Unlike (Ca,$Ln$)FeAs$_2$ whose parent compound CaFeAs$_2$ does not exist~\cite{nohara13}, EuFeAs$_2$ can be stabilized by itself~\cite{rza17}, which makes intact Eu-layers (without any substitutions) possible. The previous study~\cite{rza17} reveals that EuFeAs$_2$ undergoes two magnetic transitions at $T_\mathrm{m}^{\mathrm{Fe}}=$ 110 K and $T_\mathrm{m}^{\mathrm{Eu}}=$ 40 K, associated with spin-density-wave (SDW) order and Eu-spin antiferromagnetism, respectively. With La-doping, both transition temperatures are suppressed, and SC emerges with a maximum superconducting transition temperature ($T_\mathrm{c}$) of 11 K in Eu$_{0.85}$La$_{0.15}$FeAs$_2$. Note that this $T_\mathrm{c}$ value is much lower, compared with the highest $T_\mathrm{c}$ up to 35 K in Ca$_{0.85}$La$_{0.15}$FeAs$_2$~\cite{nohara14}, suggesting that the Eu local spins might play a negative role in raising $T_\mathrm{c}$. The decrease in $T_\mathrm{m}^{\mathrm{Eu}}$ with La doping seems to be directly related to the magnetic dilution effect. Then, it is of interest to see whether a doped EuFeAs$_2$ without disrupting the Eu-layers could be superconducting, and how the magnetism in Eu subsystem evolves with chemical doping.

In this paper, we report physical properties of Ni-doped EuFeAs$_2$ as well as the non-doped EuFeAs$_2$. The reasons for choosing Ni as the dopant are twofold: 1) Ni doping effectively induces two extra electrons in iron pnictides~\cite{cgh09,llj09}; 2) no SC can be introduced by Ni doping in 122-type EuFe$_2$As$_2$ although the SDW order is completely suppressed~\cite{rz09b}. The second point makes Ni doping more advantageous for exploring possible SC in the 112-type EuFeAs$_2$ system, because SC, once observed, should not be due to the 122-type Eu(Fe,Ni)$_2$As$_2$. As a matter of fact, we indeed observed SC with a transition temperature of 17.5 K in Eu(Fe$_{0.96}$Ni$_{0.04}$)As$_2$, which is remarkably higher than that in Eu$_{0.85}$La$_{0.15}$FeAs$_2$~\cite{rza17}. The result demonstrates that the Eu spins do not destroy SC in the 112-type system. We also found that the Eu-spin magnetism is very complicated. Below the spin-ordering temperature, there are additional magnetic anomalies which are attributed to reentrant spin-glass and spin-canting transitions. For the Ni-doped case, the spin canting is much enhanced, which generates an appreciable spontaneous magnetization that coexists with the bulk SC at low temperatures.

\section{Materials and method}\label{sec:2}

We employed a solid-state reaction in vacuum for the preparation of the polycrytalline samples of Eu(Fe$_{1-x}$Ni$_x$)As$_2$ ($x$ = 0, 0.04), similar to previous literatures~\cite{rza17,rz09b}. First, EuAs, FeAs and NiAs were prepared using source materials As (99.999\%), Eu (99.9\%), Fe (99.998\%) and Ni (99.99\%). The presynthesized arsenides were then ground into powder by agate mortar in a glove box filled with pure Ar. Second, stoichiometric mixtures of EuAs, FeAs and NiAs were homogenized, pressed into pellets, and then loaded in an alumina crucible which was sealed in a quartz ampule under 10$^{-2}$ Pa vacuum. Third, the quartz ampule was heated up to 800 $^{\circ}$C, holding for 7 days, in a muffle furnace. Finally, the sample was cooled down to room temperature. The sample was found to be stable in air.

Powder x-ray diffraction (XRD) was carried out at room temperature on a PANalytical x-ray diffractometer (Model EMPYREAN) with a monochromatic CuK$_{\alpha 1}$ radiation. The lattice parameters were refined by a least-squares fit using at least 15 XRD peaks. The electrical resistivity was measured using a standard four-electrode method, and the heat capacity was measured via a relaxation technique. Both measurements were performed on a physical property measurement system (PPMS-9, Quantum Design). The dc magnetic properties were measured on a magnetic property measurement system (MPMS-5, Quantum Design).

\section{Results and discussion}\label{sec:3}

\begin{figure}
\centering
\includegraphics[width=8cm]{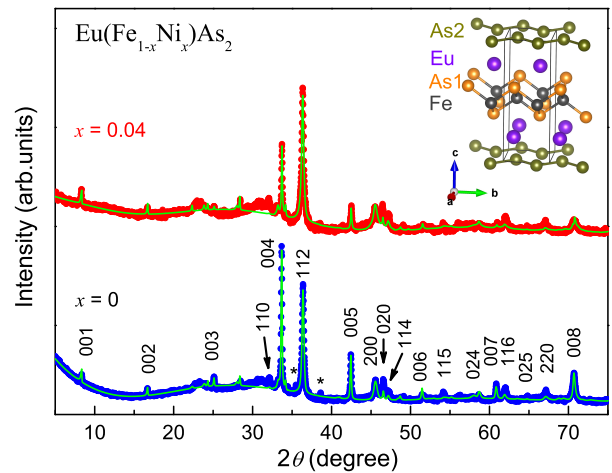}
\caption{(Color online) Powder x-ray diffraction patterns of Eu(Fe$_{1-x}$Ni$_x$)As$_2$ ($x$ = 0 and 0.04). The green lines are the calculated profiles based on the structural model in Ref.~\cite{rza17}. The small peaks labeled with asterisks come from FeAs$_2$ impurity. The inset shows the crystal structure, consisting of layers of zigzag As$_2$ chains, Eu atoms, and superconductively active Fe$_2$As$_2$ blocks.}
\label{fig1}
\end{figure}

Figure~\ref{fig1} shows the XRD patterns of the two samples Eu(Fe$_{1-x}$Ni$_x$)As$_2$ ($x$ = 0 and 0.04) studied here. First of all, the signal-to-background ratio is relatively low, compared with the case for other related materials such as Eu(Fe$_{2-x}$Ni$_x$)As$_2$~\cite{rz09b}. Similar data with weak reflections were also reported in the previous literature~\cite{rza17}. This phenomenon reflects sample's poor crystallinity, probably in relation with the relatively low sintering temperature (note that a higher sintering temperature would lead to decomposition of the 112-type phase). Note that the poor crystallinity does not influence the basic magnetic and superconducting properties measured, because the characteristic interaction lengths ($\sim$0.1-1 nm) are much shorter than the crystallite's size ($\sim$100-1000 nm). The calculated XRD profiles based on the structural model obtained from the single-crystal XRD~\cite{rza17} meet the experimental observations well, confirming that the samples are basically the 112-type phase. Nevertheless, the poor quality of the XRD data does not allow a reliable crystal-structure report.

Indeed, most of the diffraction peaks can be well indexed with a monoclinic unit cell. The refined lattice parameters of the parent compound EuFeAs$_2$ are $a$ = 3.984(1) \AA, $b$ = 3.903(1) \AA, $c$ = 10.656(2) \AA~ and $\beta$ = 89.87(2)$^{\circ}$, consistent with the previous report~\cite{rza17}. The lattice constants of the Ni-doped sample are listed in Table ~\ref{tab1}, as a comparison. The $a$ and $b$ axes of the Ni-doped sample increase by 0.08\% and 0.13\%, respectively. In contrast, the $c$ axis \emph{decreases} by 0.1\%. Overall, the change in lattice parameters resembles the case in Eu(Fe$_{2-x}$Ni$_x$)As$_2$~\cite{rz09b}, as if the lattice were uniaxially compressed along the $c$ axis. In reality, the lattice-parameter variation is related to two aspects: 1) the smaller ionic radius of Ni$^{2+}$ (compared with Fe$^{2+}$) and 2) the electron doping which tends to weaken the chemical bonding within the Fe$_2$As$_2$ layers. The decrease in $c$ is mainly due to smaller size of Ni$^{2+}$, and the increase in $b$ mostly reflects the electron doping effect.

\begin{table}
\caption{Lattice constants and magnetic/superconducting transition temperatures in Eu(Fe$_{1-x}$Ni$_x$)As$_2$ ($x$ = 0 and 0.04). $T_\mathrm{c}$, $T_\mathrm{m}^{\mathrm{Fe}}$, $T_\mathrm{m}^{\mathrm{Eu}}$, $T_\mathrm{SG}^{\mathrm{Eu}}$, and $T_\mathrm{R}^{\mathrm{Eu}}$ denote the superconducting, spin-density wave, Eu-spin ordering, possible Eu-spin-glass freezing, and possible Eu-spin canting temperatures, respectively.}
  \label{tab1}\renewcommand\arraystretch{1.3}
  \begin{tabular}{lcccccc}
      \hline \hline
  &&&EuFeAs$_2$& & &Eu(Fe$_{0.96}$Ni$_{0.04}$)As$_2$ \\
 \hline
$a$ (\r{A}) &&& 3.984(1) &&& 3.987(1)  \\
$b$ (\r{A}) & &&3.903(1) & && 3.908(1)  \\
$c$ (\r{A}) &&& 10.656(2) && &10.645(3)  \\
$\beta$ ($^{\circ}$) && &89.87(2) &&&89.77(3)  \\
    \hline
 $T_\mathrm{c}$ (K) &&& - &&& 17.5  \\
$T_\mathrm{m}^{\mathrm{Fe}}$ (K) & &&100 &&& -  \\
$T_\mathrm{m}^{\mathrm{Eu}}$ (K) &&& 40 &&& 38.5  \\
$T_\mathrm{SG}^{\mathrm{Eu}}$ (K) && &15.5& && 15.5  \\
$T_\mathrm{R}^{\mathrm{Eu}}$ (K) & &&6.5 &&& 6.2  \\
\hline\hline
  \end{tabular}
\end{table}

\begin{figure}
\centering
\includegraphics[width=7.5cm]{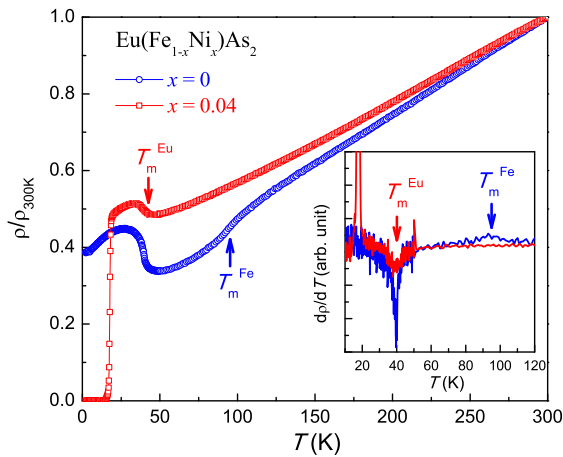}
\caption{(Color online) Temperature dependence of resistivity of the Eu(Fe$_{1-x}$Ni$_x$)As$_2$ ($x$ = 0 and 0.04) polycrystalline samples. Normalized resistivity is employed for the comparison (the absolute room-temperature resistivity is around 0.6 m$\Omega$ cm for both samples). The magnetic transitions associated with the Fe and Eu sublattices are marked with arrows. The inset plots the derivative of resistivity.}
\label{fig2}
\end{figure}

Figure~\ref{fig2} shows the temperature dependence of resistivity for the two samples of Eu(Fe$_{1-x}$Ni$_x$)As$_2$. The non-doped sample exhibits two anomalies at $\sim$40 K and $\sim$100 K, which are respectively attributed to the antiferromagnetic (AFM) orderings in the Eu and Fe sublattices. The SDW transition temperature is somewhat lower than the previous report (110 K)~\cite{rza17}, which could be due to sample's deviation from the stoichiometry (e.g. As deficiency~\cite{As-deficiency}). Also, the Eu-spin ordering at $\sim$40 K is accompanied with an obvious resistivity jump, implying the substantial coupling with the conduction electrons. For the Ni-doped sample, no sign of SDW ordering is seen. On the other hand, the Eu-spin ordering remains with a broadened transition at $\sim$40 K. SC emerges with an onset transition temperature $T_\mathrm{c}^{\mathrm{onset}}$ = 18.2 K and a midpoint temperature $T_\mathrm{c}^{\mathrm{mid}}$ = 17.6 K.

\begin{figure}
\centering
\includegraphics[width=7.5cm]{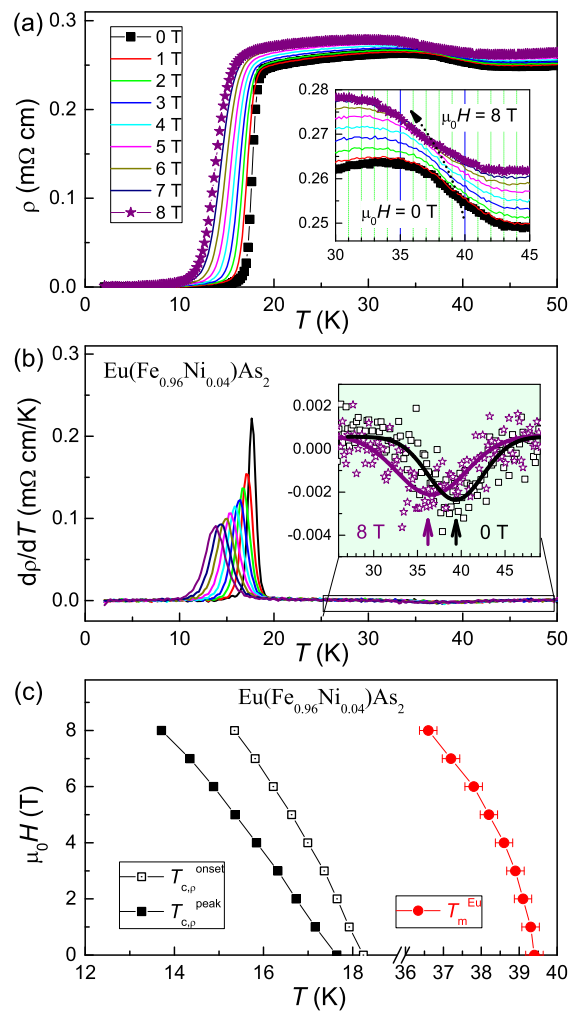}
\caption{(Color online) (a) Temperature dependence of resistivity under external magnetic fields for the Eu(Fe$_{0.96}$Ni$_{0.04}$)As$_2$ sample. The inset shows the close-up between 30 and 45 K. (b) The derivative of resistivity in (a), from which the transition temperatures are determined (the solid lines in the inset are the Gaussian fit). (c) $H-T$ phase diagram showing both the superconducting and the Eu-spin antiferromagnetic boundaries.}
\label{fig3}
\end{figure}

To investigate the magnetic field response for the superconducting and magnetic transitions in Eu(Fe$_{0.94}$Ni$_{0.04}$)As$_2$, we measured the temperature dependence of resistivity under magnetic fields up to $\mu_{0}H$ = 8 T. As shown in Fig.~\ref{fig3}(a), the superconducting transition shifts to lower temperatures with increasing the external field. Simultaneously, the Eu-spin related transition temperature also goes down, which is explicitly seen in the inset of Fig.~\ref{fig3}(b). Notably, the material shows a positive magnetoresistance with $MR=(R_H-R_0)/R_0\approx$ 5\% at $\mu_{0}H$ = 8 T, irrespective of Eu-spin ordering, in sharp contrast with the negative magnetoresistance in most electron-doped Eu-containing 122 systems~\cite{js09b,jwh13,jwh17}. This issue deserves further investigations.

Fig.~\ref{fig3}(c) displays the superconducting and magnetic phase diagram. The upper critical fields are given by using two criteria: 90\%$\rho_\mathrm{n}$ [herein $\rho_\mathrm{n}$ is the extrapolated normal-state resistivity value at $T_{\mathrm{c}}(H)$] and the peak temperature in the d$\rho$/d$T$ curves. The initial slope $\mu_{0}$d$H_{\mathrm{c2}}$/d$T$ is $-$3.13 and $-$2.27 T/K, respectively, for the two criteria. These initial slopes are over twice of those in Eu-containing 122-type ferromagnetic superconductors~\cite{rz09a,js09b}, implying negligible exchange field on the superconducting Cooper pairs in the present 112 system. The decrease in $T_\mathrm{m}^{\mathrm{Eu}}$ with $H$ suggests an \emph{antiferromagnetic} order. Nonetheless, the sensitivity of $T_\mathrm{m}^{\mathrm{Eu}}$ to external field is surprising. We will discuss on this issue later on.

\begin{figure}
\centering
\includegraphics[width=7.5cm]{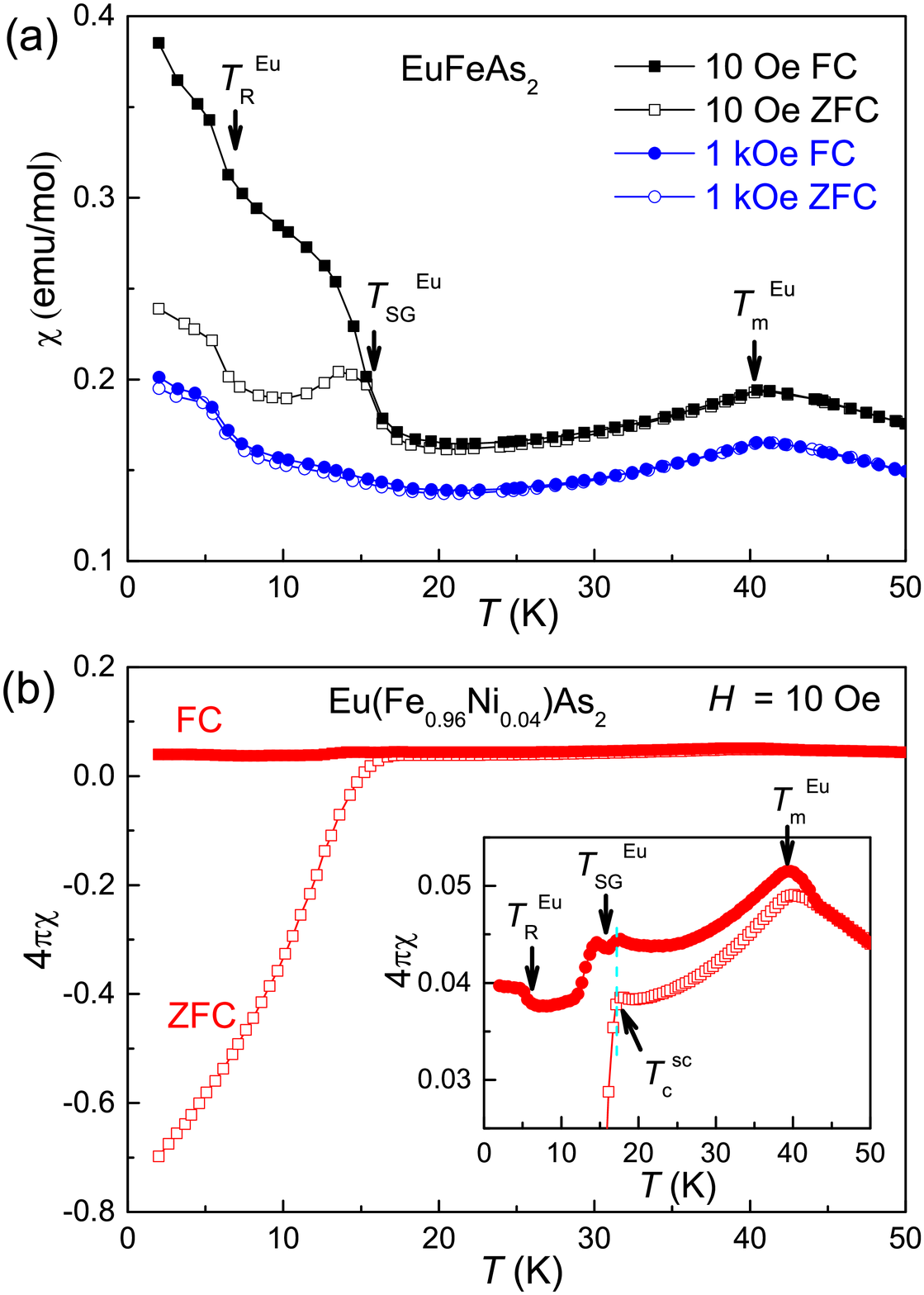}
\caption{(Color online) Temperature dependence of magnetic susceptibility of EuFeAs$_2$ (a) and Eu(Fe$_{0.96}$Ni$_{0.04}$)As$_2$ (b). The data were collected in a heating process with zero-field-cooling (ZFC) and field-cooling (FC) histories. Anomalies at $T_\mathrm{m}^{\mathrm{Eu}}$, $T_\mathrm{SG}^{\mathrm{Eu}}$, and $T_\mathrm{R}^{\mathrm{Eu}}$ are marked with arrows, possibly associated with the magnetic transitions of Eu-spin ordering, spin-glass formation, and reentrant magnetic modulation, respectively. The inset of panel (b) shows a close-up in order to highlight the superconducting transition and to reveal the successive magnetic transitions.}
\label{fig4}
\end{figure}

SC in Eu(Fe$_{0.96}$Ni$_{0.04}$)As$_2$ is confirmed by the dc magnetic susceptibility measurement shown in Fig.~\ref{fig4}(b). The onset transition temperature is 17.5 K, close to the resistive transition midpoint. The magnetic shielding volume fraction (ZFC data) at 2 K is about 70\%, suggestive of bulk SC. The FC data show only a slight decrease below $T_\mathrm{c}$. This is commonly due to flux pinning effect. Note that, in the FC data of Eu(Fe$_{0.96}$Ni$_{0.04}$)As$_2$, there are additional anomalies below $T_\mathrm{c}$. Obviously, these anomalies should come from the Eu-spin magnetism rather than SC.

\begin{figure}
\centering
\includegraphics[width=7.5cm]{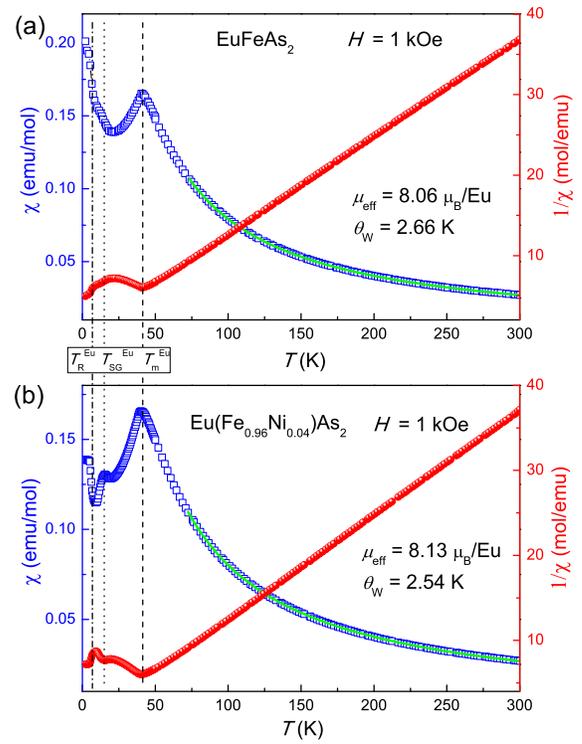}
\caption{(Color online) Temperature dependence of dc magnetic susceptibility under an external field of $H$ = 1 kOe for EuFeAs$_2$ (a) and Eu(Fe$_{0.96}$Ni$_{0.04}$)As$_2$ (b). The right axes plot the reciprocal of susceptibility. Parameters $\mu_{\mathrm{eff}}$ and $\theta_{\mathrm{W}}$ are the effective moment and the Weiss constant, respectively, obtained from the Curie-Weiss fitting (green solid lines). The dashed, dotted, and dash-dotted lines are guides to the eye for identifying the successive magnetic transitions at $T_\mathrm{m}^{\mathrm{Eu}}$, $T_\mathrm{SG}^{\mathrm{Eu}}$, and $T_\mathrm{R}^{\mathrm{Eu}}$, respectively.}
\label{fig5}
\end{figure}

It is useful to compare the magnetic data of the two samples for  unveiling the complex Eu-spin magnetism. As shown in Fig.~\ref{fig4}(a), first of all, the susceptibility decreases at 40.5 K, characteristic of an AFM ordering~\cite{rz08,blundell}. Similarly, there is an AFM-like transition at 39.5 K for Eu(Fe$_{0.96}$Ni$_{0.04}$)As$_2$ also, yet the FC and ZFC data start to separate below 43 K, reminiscent of spin-glass freezing. Secondly, the FC and ZFC curves for EuFeAs$_2$ bifurcate at $T_\mathrm{SG}^{\mathrm{Eu}}$ = 15.5 K. This feature resembles the case in EuFe$_2$(As$_{1-x}$P$_{x}$)$_2$~\cite{zapf}, suggesting a reentrant spin-glass transition. This spin-glass-like transition at $T_\mathrm{SG}^{\mathrm{Eu}}$ is accompanied with an upturn in the susceptibility, which is also detectable in Eu(Fe$_{0.96}$Ni$_{0.04}$)As$_2$ [see the inset of Fig.~\ref{fig4}(b)]. Thirdly, with further decreasing temperature, the susceptibility increases rapidly at $T_\mathrm{R}^{\mathrm{Eu}}$ = 6.5 K, below which it tends to level off. The similar behavior appears in Eu(Fe$_{0.96}$Ni$_{0.04}$)As$_2$. We speculate that it could be due to a reentrant spin canting. Future studies using single-crystal samples and various probes, such as M\"{o}ssbauer spectroscopy and neutron diffractions, will be desired in order to clarify the details of the Eu-spin ordering.

Figure~\ref{fig5} displays the dc magnetic susceptibility from 2 to 300 K under an external magnetic field of $H$ = 1 kOe. At high temperatures ($T>$ 50 K), the reciprocal of susceptibility, $1/\chi$, is essentially linear, indicating local-moment paramagnetism contributed dominantly from the Eu spins. The Curie-Weiss fitting with the formula $\chi=\chi_0+C/(T+\theta_\mathrm{W})$ yields the effective moment ($\mu_\mathrm{eff}$) and Weiss temperature ($\theta_\mathrm{W}$) for the two samples. The fitted $\mu_\mathrm{eff}$ value is very close to the theoretical one, 7.94 $\mu_{\mathrm{B}}$, of a free Eu$^{2+}$ spin, indicating essentially divalent Eu$^{2+}$ state in the Eu(Fe$_{1-x}$Ni$_x$)As$_2$ system. The Weiss temperature is positive, suggesting that AFM interactions predominate. Nevertheless, the $\theta_\mathrm{W}$ value is much lower than Eu-spin ordering temperature $T_\mathrm{m}^{\mathrm{Eu}}\sim$ 40 K. This apparent discrepancy suggests existence of both ferromagnetic and AFM interactions that are comparable in strength. The appearance of reentrant magnetic transitions below $T_\mathrm{m}^{\mathrm{Eu}}$ seems to be the consequence of the contrasting magnetic interactions.

\begin{figure}
\centering
\includegraphics[width=8cm]{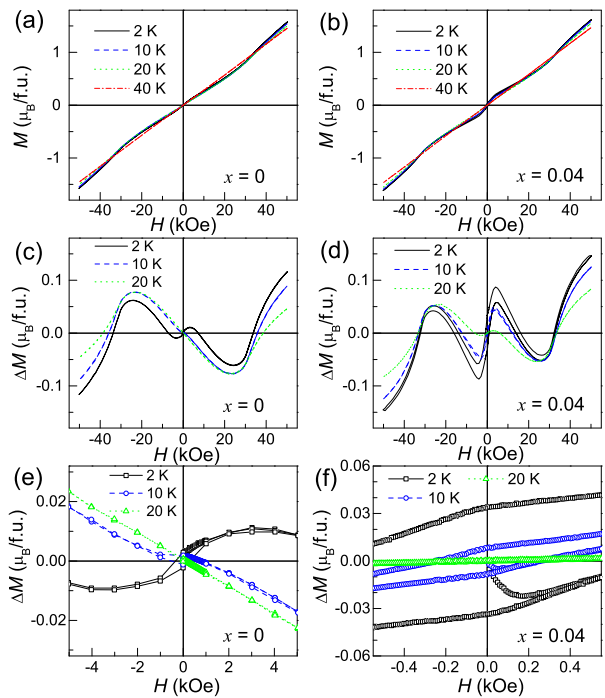}
\caption{(Color online) Isothermal magnetization of Eu(Fe$_{1-x}$Ni$_{x}$)As$_2$ with $x$ = 0 (a) and $x$ = 0.04 (b). The middle panels (c) and (d) plot the difference in magnetization, $\Delta M=M_{T}-M_{\mathrm{40K}}$, for the two samples respectively. The bottom panels (e) and (f) zoom in the $\Delta M$ data in the low-field region (note that their scales differ).}
\label{fig6}
\end{figure}

To further investigate the successive transitions, we measured the isothermal magnetization at some specific temperatures, as shown in Figs.~\ref{fig6}(a) and (b). At first sight, all the $M(H)$ data are basically linear without any tendency of saturation. The $M$ value at $\mu_0 H$ = 5 T is about 1.5 $\mu_{\mathrm{B}}$/f.u. (f.u. denotes formula unit), far below the value of 7 $\mu_{\mathrm{B}}$/f.u. expected for a parallel alignment of the Eu spins. These $M(H)$ features rule out the possibility of Eu-spin ferromagnetism that was observed in EuFe$_2$(As$_{1-x}$P$_{x}$)$_2$~\cite{rz09a,nandi14}, Eu(Fe$_{2-x}$Ni$_x$)As$_2$~\cite{rz09b}, RbEuFe$_4$As$_4$~\cite{ly16a}, and CsEuFe$_4$As$_4$~\cite{ly16b}.

On closer examination, there are subtle changes in the $M(H)$ curves. Note that $M$ is essentially proportional to $H$ at 40 K for both samples. Therefore, one may treat the data set, $M(H)_{\mathrm{40K}}$, as the reference, and the $M(H)$ curves at a fixed $T$ can be evaluated using the subtraction, $\Delta M=M_{T}-M_{\mathrm{40K}}$. The resultant $\Delta M$ curves are plotted in Figs.~\ref{fig6}(c-f). For non-doped EuFeAs$_2$, a minimum in $\Delta M$ can be seen at $H$ = 23 kOe, suggesting a metamagnetic transition. At 10 K, which is below $T_{\mathrm{SG}}^{\mathrm{Eu}}$, a small hysteresis loop appears, consistent with the spin-glass state. The hysteresis loop is enhanced, and becomes S-shape-like at 2 K, indicating existence of a ferromagnetic component. Note that the `saturation' magnetization is only $\sim$0.01 $\mu_{\mathrm{B}}$/f.u., suggesting that the ferromagnetism possibly comes from spin canting.

The $\Delta M$ curves of the Ni-doped sample share similarities on the metamagnetic transition. However, there are obvious differences. An S-shape magnetization is seen at 20 K, in contrast with the linear field dependence (within $-$10 kOe $<H<$ 10 kOe) in the non-doped EuFeAs$_2$. The S-shape magnetization is greatly enhanced at 10 and 2 K with a `saturation' magnetization close to 0.1 $\mu_{\mathrm{B}}$/f.u., possibly owing to a heavier spin canting. The spontaneous magnetization corresponds to an internal field of $\sim$100 Oe, somewhat larger than the lower critical field [see Fig.~\ref{fig6}(f)], suggesting a spontaneous vortex ground state~\cite{jwh17}. The magnetic hysteresis extends to the highest field, due to the superimposition with the superconducting loop (from flux pinning). Similar superimpositions between SC and spontaneous ferromagnetic magnetization are seen in the 122-type ferromagnetic superconductors~\cite{jwh13,jwh17b} where a full saturation magnetization of $\sim$7 $\mu_{\mathrm{B}}$/f.u. are observed.

\begin{figure}
\centering
\includegraphics[width=7.5cm]{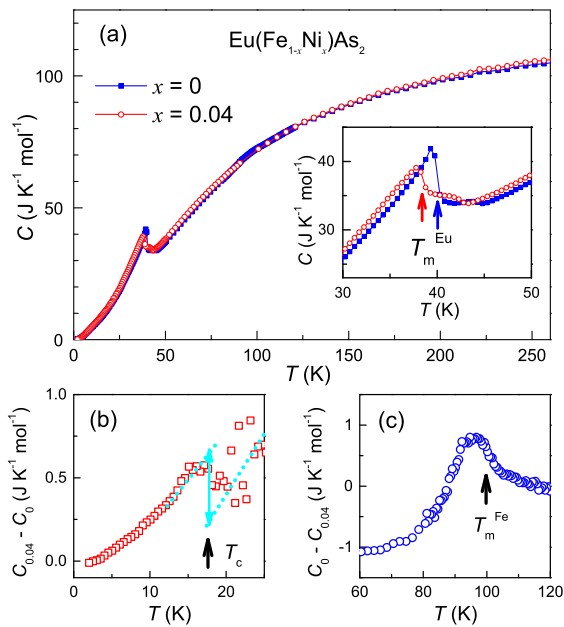}
\caption{(Color online) (a) Temperature dependence of specific heat of Eu(Fe$_{1-x}$Ni$_{x}$)As$_2$ ($x$ = 0 and 0.04). The inset is a close-up for the anomaly at around 40 K. The bottom panels (b) and (c) plot the differences of specific heat, $C_{x=0.04}-C_{x=0}$ and $C_{x=0}-C_{x=0.04}$, respectively. The superconducting and SDW transitions are marked with arrows.}
\label{fig7}
\end{figure}

Figure~\ref{fig7}(a) shows the temperature dependence of specific heat for Eu(Fe$_{1-x}$Ni$_{x}$)As$_2$ ($x$ = 0 and 0.04). Prominent anomalies at around 40 K can be seen immediately, which are attributed to the Eu-spin ordering. For the $x=0$ sample, a large specific-heat jump appears at $T_\mathrm{m}^{\mathrm{Eu}}=$ 40 K, and an additional slight anomaly at 44.4 K can be detected. The main jump is weakened with a reduced transition temperature of $T_\mathrm{m}^{\mathrm{Eu}}=$ 38.5 K for the Ni-doped sample. Meanwhile, the secondary transition at 42.5 K is enhanced. Combining with the above resistivity and magnetic measurement results altogether, we conclude that most of the Eu spins order at $T_\mathrm{m}^{\mathrm{Eu}}$. The second transition, $\sim$4 K above above $T_\mathrm{m}^{\mathrm{Eu}}$, is coincident with the magnetic-susceptibility bifurcation shown in Fig.~\ref{fig4}(b). Thus, it is likely to be intrinsic (rather than phase separations). The transition could be a static short-range magnetic ordering (like a cluster-spin glass), serving as the precursor of AFM ordering. Note that there are no detectable signals at $T_\mathrm{SG}^{\mathrm{Eu}}$ and $T_\mathrm{R}^{\mathrm{Eu}}$, corresponding to the possible reentrant spin-glass and spin-canting transitions, respectively. This is not surprising since reentrant magnetic transitions generally do not cause an obvious change in entropy.

To look for the expected specific-heat anomalies at the SDW and superconducting transitions, we made a specific-heat subtraction ($C_{x=0.04}-C_{x=0}$ and $C_{x=0}-C_{x=0.04}$) between the two samples. As shown in Fig.~\ref{fig7}(c), a lambda-shape anomaly is seen at around 100 K, confirming the SDW ordering for the non-doped EuFeAs$_2$. Meanwhile, a specific-heat jump is revealed at $T_\mathrm{c}=$ 17.5 K. The amplitude of the jump is roughly estimated to be $\sim$0.4 J K$^{-1}$ mol$^{-1}$. The Sommerfeld constant $\gamma_\mathrm{n}$ is difficult to be extracted here, yet it should be close to those of the electron-doped iron-based superconductors with a similar $T_\mathrm{c}$ value, e.g. 17 mJ K$^{-2}$ mol$^{-1}$ for BaFe$_{1.9}$Ni$_{0.1}$As$_2$ ($T_\mathrm{c}$ = 20.1 K)~\cite{lhq10} and 17.2 mJ K$^{-2}$ mol$^{-1}$ for BaFe$_{1.9}$Co$_{0.1}$As$_2$ ($T_\mathrm{c}$ = 19.5 K)~\cite{hardy10}. Therefore, the dimensionless parameter $\Delta C/(\gamma_\mathrm{n} T_\mathrm{c})$ is around 1.3, close to the value of BCS weak-coupling scenario (1.43), confirming bulk SC in Eu(Fe$_{0.96}$Ni$_{0.04}$)As$_2$.

Now let us discuss the Eu magnetism in Eu(Fe$_{1-x}$Ni$_{x}$)As$_2$. Because the Eu$-$Eu interatomic distance is about 4.0 \AA, larger than the Hill limit (3.5 \AA), no appreciable direct exchange interactions between the Eu spins are expected. Therefore, the magnetic interactions should be dominated by indirect interactions such as superexchange via As-4$p$ orbitals and the Ruderman-Kittel-Kasuya-Yosida (RKKY) exchange via conduction electrons. In particular, the RKKY coupling strength, $J_{\mathrm{RKKY}}$, is proportional to $\mathrm{cos}(2k_{\mathrm{F}}r)/r^{3}$, where $k_\mathrm{F}$ is the Fermi vector and $r$ is the distance between the local moments, in a simplified scenario~\cite{blundell}. That is to say, $J_{\mathrm{RKKY}}$ is of long-range and, with both ferromagnetic and AFM interactions. Furthermore, there are two asymmetric Eu$^{2+}$-layers in the unit cell of Eu(Fe$_{1-x}$Ni$_{x}$)As$_2$, which give rise to additional geometrical frustrations. Taken all these factors together, it is not surprising for the complex magnetism of Eu spins in Eu(Fe$_{1-x}$Ni$_{x}$)As$_2$. Besides, the sensitivity of $T_\mathrm{m}^{\mathrm{Eu}}$ to external magnetic fields might also reflect the multiple magnetic interactions among the local spins on an asymmetric lattice.

Since EuFeAs$_2$ is the unique 112-type parent compound, the Ni-doping-induced SC reported here represents the first example for 112 systems to achieve SC by merely Fe-site doping. Note that the maximum $T_\mathrm{c}$ in Eu$_{0.85}$La$_{0.15}$FeAs$_2$ is 11 K (the onset diamagnetic transition temperature is actually about 8 K)~\cite{rza17}, in this context, the $T_\mathrm{c}$ value of 17.5 K in Eu(Fe$_{0.96}$Ni$_{0.04}$)As$_2$ is rather impressive. Assuming that each Ni atom induces two extra electrons~\cite{cgh09,llj09}, the electron-doping level is only 8\%, obviously lower than that in Eu$_{0.85}$La$_{0.15}$FeAs$_2$. Therefore, the carrier-doping level cannot account for the $T_\mathrm{c}$ difference. In fact, the SDW order is not completely suppressed in Eu$_{0.85}$La$_{0.15}$FeAs$_2$~\cite{rza17}, which could be the main reason for the low $T_\mathrm{c}$. The Fe-site electron doping in Eu(Fe$_{1-x}$Ni$_{x}$)As$_2$ seems to suppress the SDW order more efficiently, thus giving rise to a relatively high $T_\mathrm{c}$.

\section{Conclusion}\label{sec:4}

In summary, we have synthesized and characterized Eu-containing 112-type iron arsenides EuFeAs$_2$ and Eu(Fe$_{0.96}$Ni$_{0.04}$)As$_2$, both of which were systematically studied with the transport, magnetic, and thermodynamic measurements. EuFeAs$_2$ was found to undergo an SDW-like transition at $\sim$100 K, whilst Eu(Fe$_{0.96}$Ni$_{0.04}$)As$_2$ was discovered to show a bulk SC at 17.5 K without evidence of any SDW-like order. Both materials exhibit an Eu-spin AFM transition at $\sim$40 K, followed by possible reentrant spin-glass and spin canting transitions. Notably, Eu(Fe$_{0.96}$Ni$_{0.04}$)As$_2$ displays an enhanced spontaneous magnetization that coexists with SC. Thus, the present Eu(Fe$_{1-x}$Ni$_{x}$)As$_2$ system provides an additional playground for studying the influence of internal field on superconductivity.

\begin{acknowledgments}
This work was supported by the National Natural Science Foundation of China (Grant No. 11474252), the National Key Research and Development Program of China (No.2016YFA0300202), and the Fundamental Research Funds for the Central Universities of China.
\end{acknowledgments}

\end{document}